\documentstyle[12pt,twoside,fleqn,espcrc1,epsfig]{article}
% ...

\newcommand{\AmS}{{\protect\the\textfont2
  A\kern-.1667em\lower.5ex\hbox{M}\kern-.125emS}}
 
\title{\large{Isotope correlations as a probe for freeze-out characterization:
central $^{124}$Sn+$^{64}$Ni, $^{112}$Sn+$^{58}$Ni collisions }}

\author{E. Geraci\address{\vspace{-0.27cm} \baselineskip=1pt\small INFN and 
Dipartimento di Fisica, Universit\`a di Bologna, Italy}$^{,}$
\address{\vspace{-0.27cm} \baselineskip=1pt\small INFN-LNS and Dipartimento 
di Fisica e Astronomia, Universit\`a di Catania, Italy}, 
M.~Alderighi\address{\vspace{-0.27cm} \baselineskip=1pt\small INFN and Istituto di 
Fisica Cosmica, CNR, Milano, Italy},
 A.~Anzalone$^{b}$, L.~Auditore\address{\vspace{-0.27cm} 
\baselineskip=1pt\small INFN and Dipartimento di Fisica, Universit\`a di Messina, 
Italy} V.~Baran$^{b,}$\address{\vspace{-0.27cm} \baselineskip=1pt\small 
Institute for Physics and Nuclear Engineering, Bucharest, Romania},
M.~Bartolucci\address{\vspace{-0.27cm} \baselineskip=1pt\small INFN and Dipartimento 
di Fisica, Universit\`a di Milano, Italy}, 
I.~Berceanu$^{e}$, J.~Blicharska\address{\vspace{-0.27cm} 
\baselineskip=1pt\small Institute of Physics, University of Silesia, Katowice, 
Poland}, A.~Bonasera$^{b}$, B.~Borderie\address{\vspace{-0.27cm} 
\baselineskip=1pt\small Institut de Physique nucl\'eaire,
 IN2P3-CNRS, Orsay, France}, R.~Bougault\address{\vspace{-0.27cm} 
\baselineskip=1pt\small LPC, ENSI Caen 
and Universit\'e de Caen, France}, M.~Bruno$^{a}$, 
J.~Brzychczyk\address{\vspace{-0.27cm} \baselineskip=1pt\small 
M.~Smoluchowski Institute of Physics,Jagellonian 
University, Cracow, Poland}, G.~Cardella\address{\vspace{-0.27cm} 
\baselineskip=1pt\small INFN and Dipartimento di Fisica e Astronomia, 
Universit\`a di Catania, Italy}, S.~Cavallaro$^{b}$, 
A.~Chbihi\address{\vspace{-0.27cm} \baselineskip=1pt\small GANIL, CEA, 
IN2P3-CNRS, Caen, France}, J.~Cibor\address{\vspace{-0.27cm} 
\baselineskip=1pt\small H. Niewodniczanski Institute of Nuclear 
Physics, Cracov, Poland}, M.~Colonna$^{b}$, M.~D'Agostino$^{a}$, 
E~.De~Filippo$^{k}$, M.~Di~Toro$^{b}$, F.~Giustolisi$^{b}$, A.~Grzeszczuk$^{g}$,
P.~Guazzoni$^{f}$, D.~Guinet\address{\vspace{-0.27cm} \baselineskip=1pt\small 
Institut de Physique nucl\'eaire, IN2P3-CNRS, Lyon, France}, 
M.~Iacono-Manno$^{b}$, S.~Kowalski$^{g}$, 
E.~La~Guidara$^{b}$, G.~Lanzalone$^{b}$, G.~Lanzan\'{o}$^{k}$, 
N.~Le~Neindre$^{h}$, S.~Li\address{\vspace{-0.27cm} 
\baselineskip=1pt\small Institute of Modern Physics Lanzhou,
 China}, S.~Lo~Nigro$^{k}$, C.~Maiolino$^{b}$, Z.~Majka$^{j}$, 
G.~Manfredi$^{f}$, T.~Paduszynski$^{g}$, A.~Pagano$^{k}$, M.~Papa$^{k}$, 
M.~Petrovici$^{e}$, E.~Piasecki\address{\vspace{-0.27cm} 
\baselineskip=1pt\small Institute of Experimental 
Physics, University of Warsaw, Poland}, S.~Pirrone$^{k}$, G.~Politi$^{k}$, 
A.~Pop$^{e}$, F.~Porto$^{b}$, M.~F.~Rivet$^{h}$, 
E.~Rosato\address{\vspace{-0.27cm} \baselineskip=1pt\small INFN and Dipartimento di 
Fisica, Universit\`a di Napoli, Italy}, S.~Russo$^{f}$, P.~Russotto$^{b}$, 
G.~Sechi$^{c}$, V.~Simion$^{e}$, M.~L.~Sperduto$^{b}$, J.~C.~Steckmeyer$^{i}$,  
A.~Trifir\`o$^{d}$, M.~Trimarchi$^{d}$, G.~Vannini$^{a}$, M.~Vigilante$^{q}$, 
J.~P.~Wieleczko$^{l}$, J.~Wilczynski\address{\vspace{-0.27cm} 
\baselineskip=1pt\small Institute for Nuclear Studies, Otwock-Swierk, Poland},
 H.~Wu$^{o}$, Z.~Xiao$^{o}$, L.~Zetta$^{f}$,
 W.~Zipper$^{g}$
}  
 
\begin{document}
% typeset front matter
\maketitle

\begin{abstract}
$^{124}$Sn+$^{64}$Ni and $^{112}$Sn+$^{58}$Ni reactions at 35 AMeV
 incident energy were studied with the forward part of CHIMERA multi-detector.
 The most central collisions  were selected by means of a multidimensional 
analysis.
The characteristics of the source formed in the central collisions, 
as size, temperature and volume, were inspected.
The measured isotopes of light fragments
($3 \le Z \le 8$) were used to examine isotope yield ratios that provide
 information on the free neutron to proton densities.
\end{abstract}
 
\section{INTRODUCTION}
The isotopic composition of nuclear reaction products provides important 
information on the reaction dynamics and the possible occurrence of a 
liquid-gas phase transition~\cite{isobook}.
Isotopically resolved light fragments have revealed systematic
scalings~\cite{isoscaling} that supply information on neutron and proton 
densities at the breakup stage. 
The isoscaling has been observed in a variety of reactions under the 
assumptions of statistical emissions and equal temperature and volume of the 
two considered systems. 
Our study was devoted to select central collisions, inspect the 
characteristics of the source formed in the central events (size, temperature 
and volume), and examine the isotopic ratios. 
In particular, using fragments with a charge $3\le Z \le8$ identified in mass,
we studied the isoscaling and the isotopic ratios for mirror nuclei.
We present data concerning $^{124}$Sn+$^{64}$Ni
and $^{112}$Sn+$^{58}$Ni central collisions at 35 AMeV.
The measurements have been performed using the 688 Si-CsI telescopes 
of the forward part ($1^o \le \theta_{lab} \le 30^o$) of 
CHIMERA apparatus in the framework of Reverse Experiments~\cite{pagano}.

\section{EVENT SELECTION AND SOURCE CHARACTERIZATION}
In order to have an impact parameter selector we resorted 
to the principal component analysis (PCA), which is a multivariate analysis 
technique~\cite{pca} concerned with determination of the so-called principal 
variables, linear combinations of the primary physical variables.
In our case, the first three principal components (P1, P2, P3) define 
principal planes where the data retain about 80\% of the original information.
The experimental events, projected on the P1-P2 plane, show three well 
distinct clouds. Two bumps correspond to events that can be associated to 
peripheral collisions. The third cloud mostly contains well 
detected multifragmentation events, characterized by high IMF multiplicity.
A variable summarizing both the information provided 
by P1 and P3 allows for a more careful selection in order to separate 
central and semi-peripheral collisions and to select events with an isotropic
 distribution of $cos(\Theta_{flow})$.

\begin{figure} [ht]
\begin{center}
\vspace{-0.8cm}
\mbox{
\epsfig{file=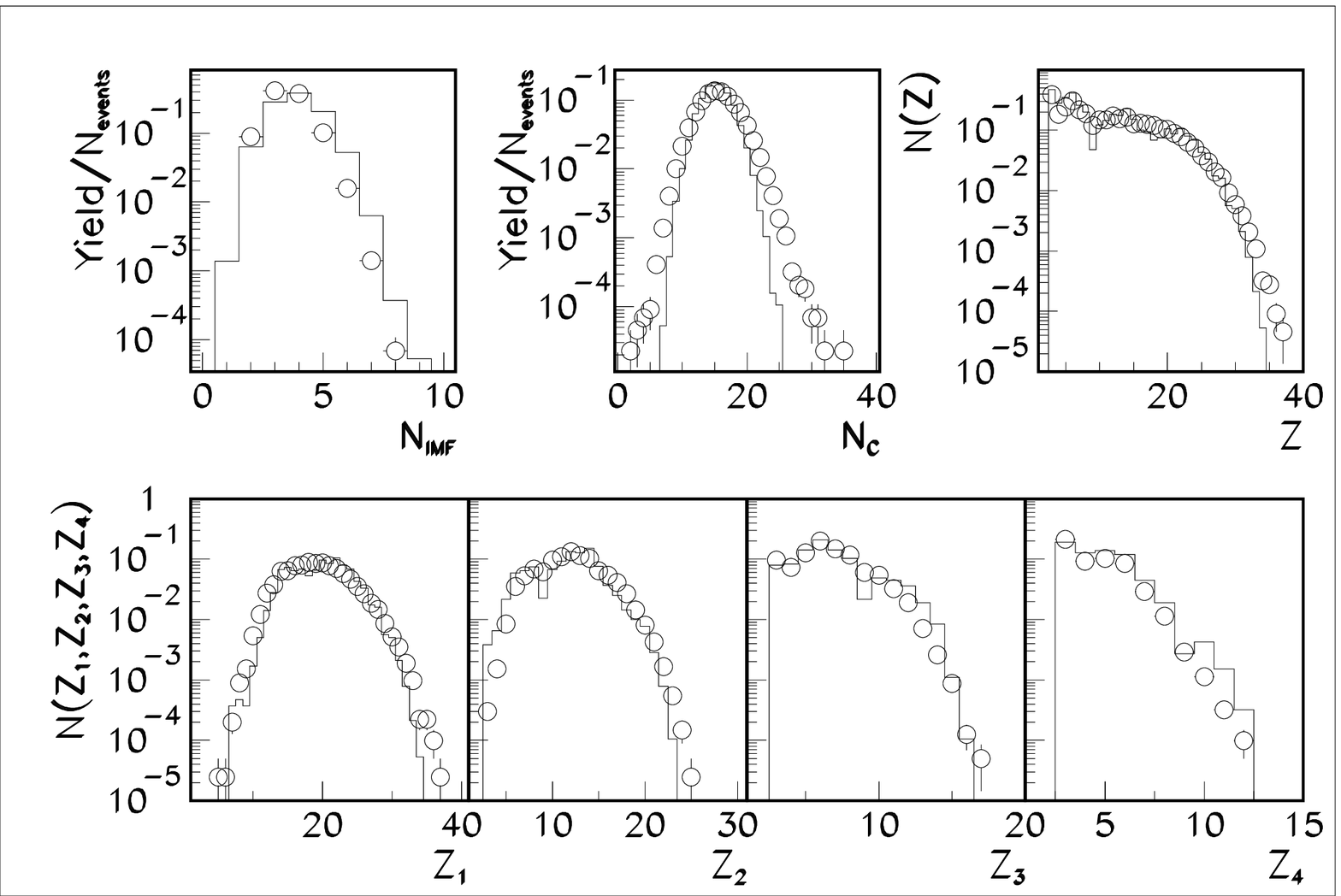,height=6.8cm}
\epsfig{file=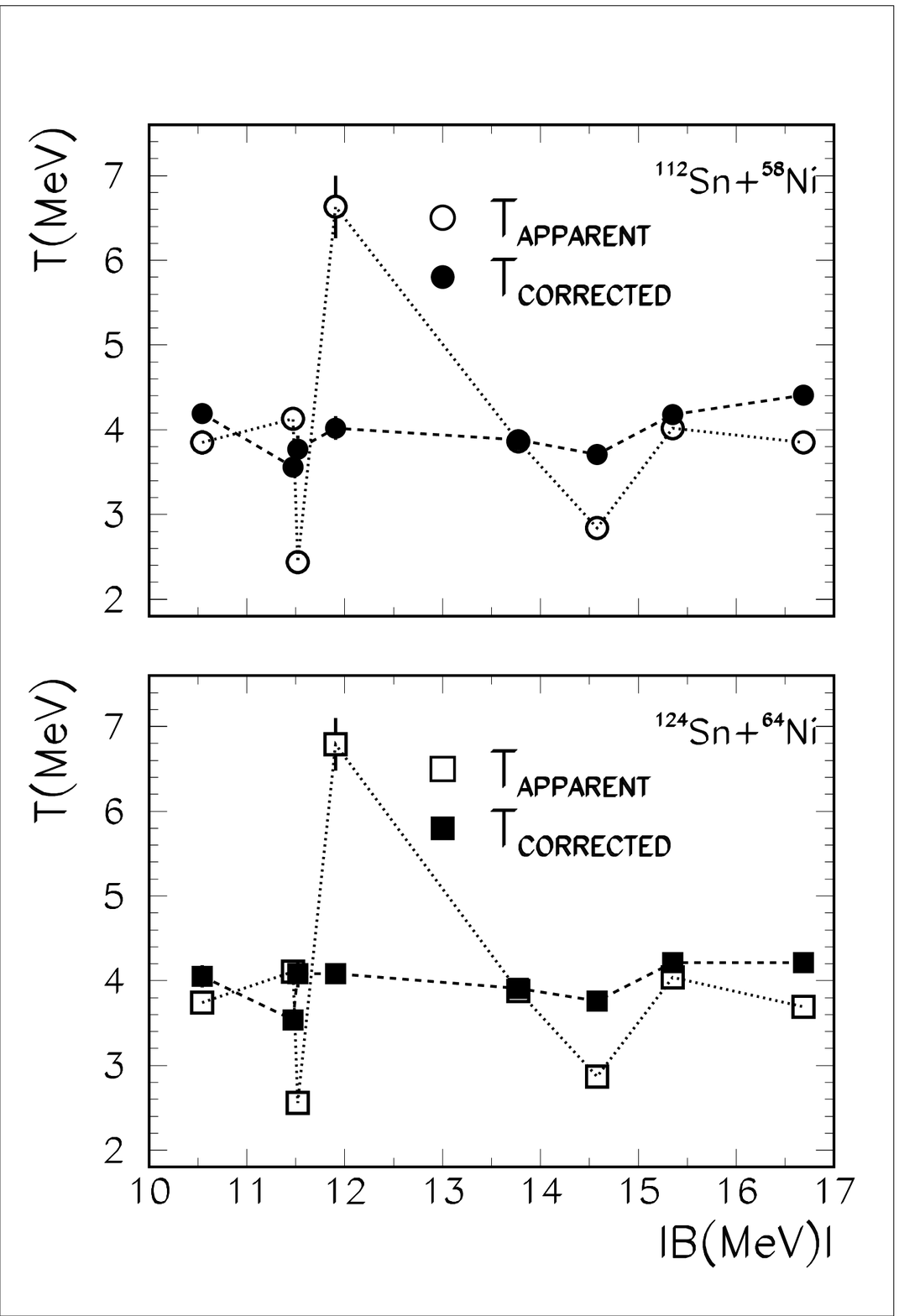,height=6.8cm}
}
\vspace{-1.cm}
\caption{\label{smm}\baselineskip=1pt\small Left box: IMF multiplicity, total 
charged multiplicity, charge distribution and charge partition in each event 
($Z_{1} \ge Z_{2} \ge Z_{3} \ge Z_{4}$).
Circles represent data, lines SMM filtered predictions for central 
$^{112}$Sn+$^{58}$Ni collisions.
Right box: The apparent and corrected temperatures as a function of the 
difference in binding energies for the $^{112}$Sn+$^{58}$Ni (top) and  
$^{124}$Sn+$^{64}$Ni (bottom) systems.
}
\end{center}
\vspace{-1.4cm}
\end{figure}

To get information on the characteristics of the source, formed in central 
collisions, we performed a comparison between data and statistical model 
(SMM~\cite{smm}) predictions.
The statistical model reproduces quite well the experimental observables 
(Fig.~\ref{smm}) if, for n-poor (rich) reaction, a source is considered having 
mass $A=145 (160)$, charge $Z=66$, excitation energy $5 AMeV$ and a freeze-out 
volume of $3V_{0}$. 
These source characteristics are in good agreement with the predictions of the 
dynamical model BNV~\cite{bnv}, which predicts
 for the n-poor (rich) reaction that an equilibrated 
excited system with $A=150 (164)$, $Z=67 (68)$ and $E^{\ast}/A =4.7 (5.1)$ MeV
is formed at a time 120-140 fm/c after the initial stage of the collision. 

Information about the space-time evolution of the reaction zone has been 
obtained via intensity interferometry. The two-fragments velocity correlation 
function for IMFs ($Z\ge3)$ reveals that the sources formed in 
$^{124}$Sn+$^{64}$Ni and $^{112}$Sn+$^{58}$Ni central collisions decay by 
emitting fragments and isotopes showing almost identical space-time 
patterns~\cite{geraci}.
 
The temperatures of the two sources were extracted by looking at the double 
isotope ratios thermometers~\cite{albergo}. The temperature is given by 
 $T=\frac{B}{ln(sR)}$ 
where s is related to the ground state spins, R is the ratio of the production
 yields of four isotopes and $B$ is related to the isotope binding energies. 
The measured yields of isotopes from Lithium to Oxygen were used to construct 
the so-called apparent temperatures; their values and the values after secondary
 decay correction~\cite{temp_corr} are presented as a function of the binding 
energy term B in the right box of Fig.~\ref{smm}. 
The deduced values of the temperatures are in the range typical
for reaction processes near the onset of multi-fragment
emissions~\cite{temp_corr,temp}.
The weighted average corrected temperatures result $3.96 \pm  0.04$ and 
$3.94 \pm 0.03$MeV for $^{112}$Sn+$^{58}$Ni and $^{124}$Sn+$^{64}$Ni reactions
respectively.  

\section{ISOTOPES ANALYSIS}
\begin{figure} [t]
\begin{center}
\vspace{-0.8cm}
\mbox{
\hspace{-0.5cm}
\epsfig{file=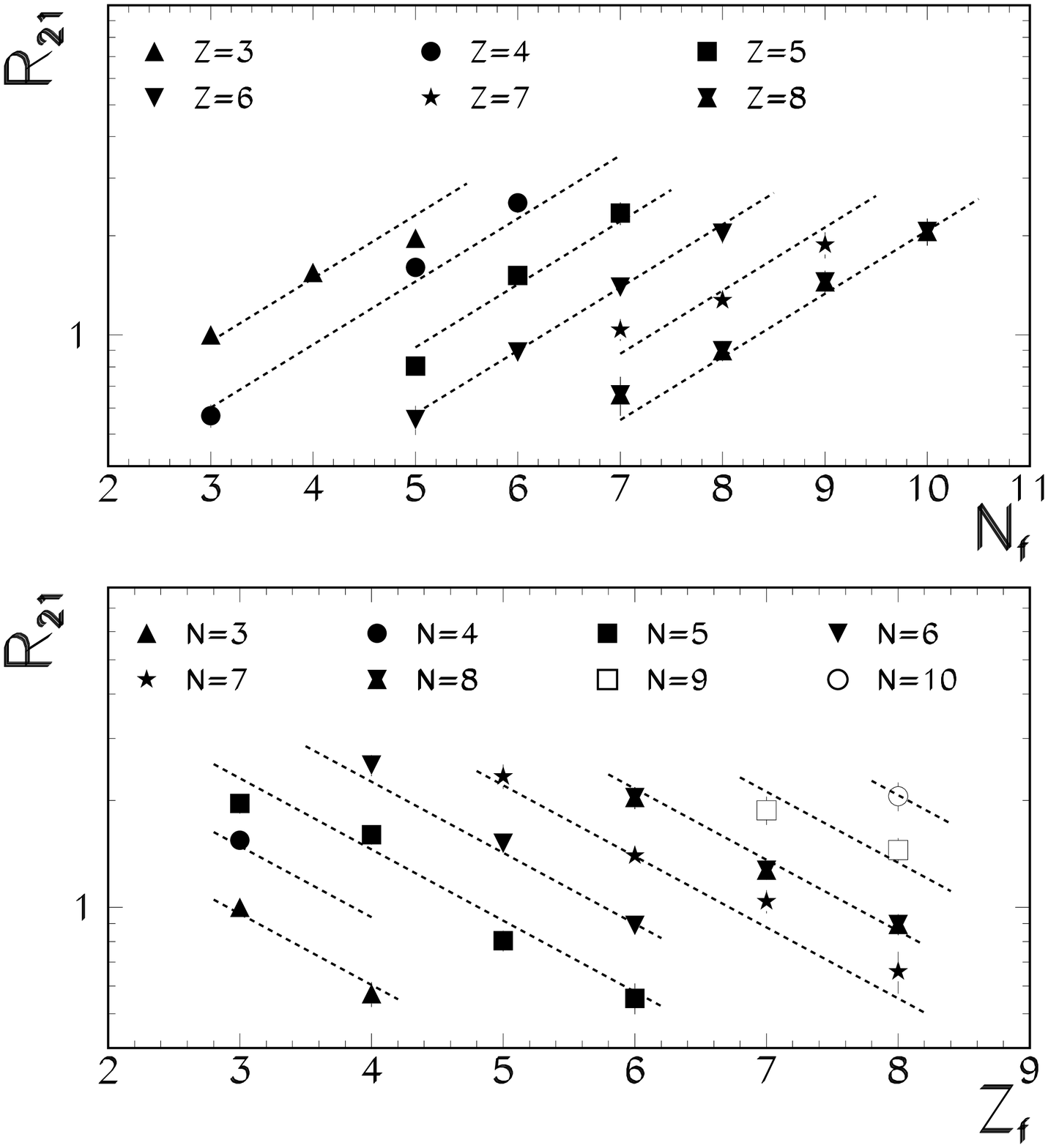,height=8.cm}
\hspace{-1.5cm}
\epsfig{file=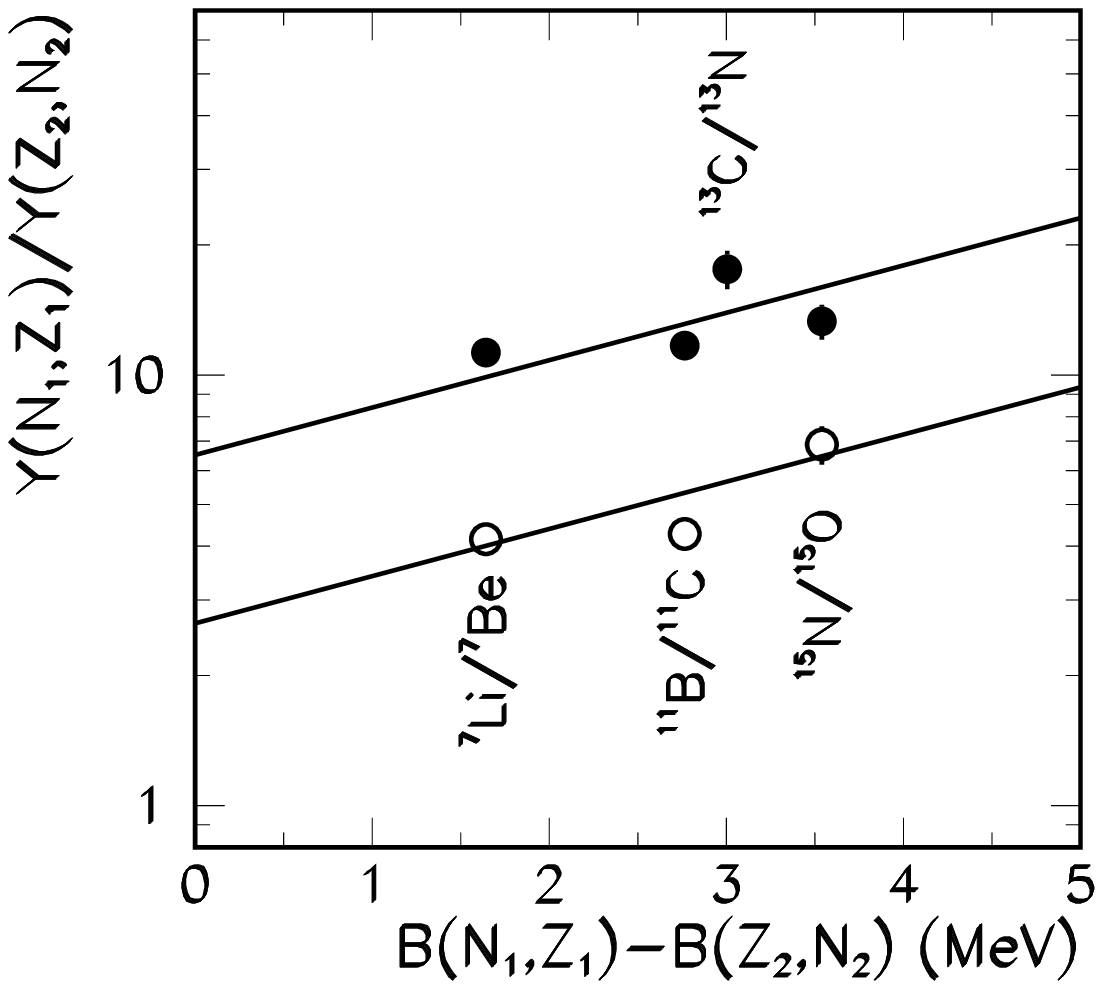,height=8.cm}}
\vspace{-1.7cm}
\caption{\label{isoscaling} \baselineskip=1pt\small 
Left panels: Isotopic ratio $R_{21}$ versus the neutron number 
$N_{f}$ (top) and the proton number $Z_{f}$ (bottom) of 
detected fragments. Right panel: Isobar ratios as a function of 
$\Delta$B for mirror nuclei in $^{112}$Sn+$^{58}$Ni (open circles)
 and $^{124}$Sn+$^{64}$Ni (solid points) reactions. Lines represent the result 
of the constrained fit.}
\end{center}
\vspace{-1.1cm}
\end{figure}

Having verified that the sources of the two analysed systems present the same 
size, temperature, volume and emission patterns, we moved on to study the 
isotopic ratios. In the Grand-Canonical 
approximation the yield ratio $R_ {21}$ of a fragment of N neutrons and Z 
protons, emitted in two reactions differing only in isospin, is related to 
the relative free neutron density and to the relative free proton density by 
the relation (neglecting the particle unstable feeding corrections~\cite{iso_decay}): \\ 
$R_{21}(N,Z)=\frac{Y_{^{124}Sn+^{64}Ni}(N,Z)}{Y_{^{112}Sn+^{58}Ni}(N,Z)} =
 C (\frac{\rho_{n,2}}{\rho_{n,1}})^N (\frac{\rho_{p,2}}{\rho_{p,1}})^Z
= C \hat{\rho_n}^N\hat{\rho_p}^Z.$\\
C is an overall normalization factor and $\hat{\rho_n}$,
$\hat{\rho_p}$ are the relative free neutron and proton densities,
related to the isoscaling parameters $\alpha$ and $\beta$~\cite{isoscaling}.
In Fig.~\ref{isoscaling}  the measured isotopic ratios are plotted as a 
function of neutron number (left upper panel) and proton number 
(left lower panel) of fragments. On the right panel of Fig.~\ref{isoscaling} 
the mirror isobar ratios are shown. 
If the sequential decay and the Coulomb effects are small, the
dependence of isobaric mirror yield ratios on the binding energy
difference should be exponential, i.e. of the form 
$(\rho_n/\rho_p)_i\exp(\Delta B/T_i)$ 
where $T_i$ is the temperature of the emitting source for the reaction
$i$ and $(\rho_n)_i$ and $(\rho_p)_i$ are the free neutron and proton
densities for the same reaction.
A constrained fit on the 27 experimental ratios of Fig.~\ref{isoscaling}
has been performed. The average corrected isotope temperature is assumed as the 
temperature of the emitting systems.
The relative neutron and proton densities  $\hat{\rho_n}$, $\hat{\rho_p}$, C 
and the free neutron to proton density $(\rho_n/\rho_p)_{124}$ (for the
reaction $^{124}$Sn + $^{64}$Ni) were used as free parameters of the fit.
The $(\rho_n/\rho_p)$ for the reaction $^{112}$Sn +$^{58}$Ni 
has been derived as $(\rho_n/\rho_p)_{112} = (\rho_n/\rho_p)_{124} \cdot
\hat{\rho_p}/\hat{\rho_n}$.
A very good agreement with an exponential 
behaviour has been obtained giving values $\hat{\rho_n}=1.55 \pm 0.02$, 
$\hat{\rho_p}=0.63\pm 0.01$ (corresponding to isoscaling parameters 
$\alpha=0.44\pm0.01$ and $\beta=-0.46\pm 0.02$), 
$(\rho_n/\rho_p)_{124}= 6.5 \pm 0.2$ and $(\rho_n/\rho_p)_{112}= 2.6 \pm 0.1$.

It can be observed that the extracted values of $\hat{\rho_n}$ and 
$(\rho_n/\rho_p)$ result definitely higher than the values calculated (1.08,
1.41 and 1.18) assuming neutrons and protons homogeneously distributed in 
a volume proportional to the nucleons number, supporting the idea of the 
isospin distillation mechanism.
Even if this can be a signal expected for a liquid-gas phase transition 
further analyses are required to evaluate secondary decay effects.

\end{document}